\documentclass{elsart}

\usepackage{graphicx}
\usepackage{epsfig,rotating}
\usepackage[dvips]{color}

\def\be{\begin{equation}}
\def\ee{\end{equation}}
\def\ba{\begin{eqnarray}}
\def\ea{\end{eqnarray}}

\def\lsim{\raise0.3ex\hbox{$\;<$\kern-0.75em\raise-1.1ex\hbox{$\sim\;$}}}
\def\gsim{\raise0.3ex\hbox{$\;>$\kern-0.75em\raise-1.1ex\hbox{$\sim\;$}}}

\def\theta{\vartheta}

\begin{document}

\begin{frontmatter}
\title{Ultra-high energy cosmic rays from a finite number of point sources}

\author[MPI]{M.~Kachelrie\ss}
\author[UCLA,INR]{and D.~Semikoz}

\address[MPI]{Max-Planck-Institut f\"ur Physik
(Werner-Heisenberg-Institut),
F\"ohringer Ring 6, 80805 M\"unchen, Germany}
\address[UCLA]{Department of Physics and Astronomy, UCLA, Los
  Angeles, CA 90095-1547} 
\address[INR]{INR RAS, 60th October Anniversary prospect 7a,
  117312 Moscow, Russia}


\begin{abstract}
We have calculated the probability that the clustering of arrival 
directions of ultra-high energy cosmic rays (UHECRs) is consistent with a
finite number of uniformly distributed proton sources, assuming the
case of small deflections by magnetic fields outside the Galaxy. A
continuous source distribution is mimicked only by an unrealisticly
high source density, $n_s\gg 10^{-2}/{\rm Mpc}^3$. Even for
densities as large as $n_s=10^{-3}/{\rm Mpc}^3$, less than half of 
the observed cluster are on average by chance. For the best-fit value
$n_s=(1$--$4)\times 10^{-5}/{\rm Mpc}^3$ derived from the AGASA data,
the probability that at least one observed cluster is from a true
point source is larger than $99.97\%$, while on average almost all
observed clusters are true. The best-fit value found is
comparable to the density of AGNs and consistent with the recent HiRes
stereo data. In this scenario, the Pierre Auger
Observatory will not only establish the clustering of UHECRs but also 
determine the density of UHECR sources within a factor of a few  after
one year of data taking.

\begin{small}
\end{small}
\end{abstract}
\end{frontmatter}

\section{Introduction}
The acceleration of protons or heavy nuclei to energies $E>10^{19}$~eV
is difficult for all known astrophysical sources of cosmic
rays~\cite{acc}. Therefore, one expects that only a small fraction of
all cosmic ray (CR) sources is able to accelerate beyond
$E>10^{19}$~eV. The signature of a small number of ultra-high energy
(UHE) CR sources is the small-scale clustering of their arrival
directions, if the deflection of CRs in magnetic fields
can be neglected. The structure and magnitude of the Galactic magnetic
field can be estimated observing the Faraday rotation of the
polarized radio-emission of pulsars. 
At energies $E>4\times10^{19}$~eV, the deflection of CR
protons in this field is less than 4--6 degrees in most directions and
decreases to 1--2 degrees at $E>10^{20}$~eV~\cite{gmf}. The overall
effect on potential CR clusters is even smaller when the energies of the
CRs in the cluster are not too different.

The magnitude and structure of extragalactic magnetic fields is more
uncertain. Only recently, magnetic fields were included in simulations
of large scale structures~\cite{Dolag:2003ra,ems}. It was found that
extragalactic magnetic fields are strongly localized in galaxy
clusters and filaments, while voids contain only primordial
fields. The latter cannot be stronger than $B\sim 10^{-11}$~G,
otherwise the observed field strengths in galaxy clusters would be
exceeded. Even if sources tend to sit in regions of high density and
thus strong magnetic fields, CRs can be significantly deflected only
within clusters. But the angular size of distant galaxy clusters is much
less than one degree and thus they appear as point-like
sources unless a nearby cluster is on the line of sight. 
Thus in the part of the sky outside of nearby galaxy clusters 
astronomy with UHE protons may be possible, unless the observer is
embedded in a strongly magnetized region. In the latter case,
deflections are important even for UHE protons, and charged particle
astronomy may not be possible~\cite{ems}. A crucial step towards 
the goal of UHE proton astronomy is the identification of point
sources of UHECRs.

The AGASA data on the arrival direction of CRs with energies $E>4\times
10^{19}$~eV contain a clustered component with five pairs and one
triplet within $2.5$
degrees~\cite{AGASAcluster_data,AGASAcluster_data2}. 
Neglecting possible systematic errors in energy scales, the
sensitivity of the other experiments for clustering at the energies 
$E>4 \times 10^{19}$~eV is much smaller, either because of the
smaller exposure at the highest energies (Yakutsk, HiRes in stereo
mode) or because of a poor two-dimensional angular resolution (HiRes
in monocular mode). 
At lower energies, $E < 4 \times 10^{19}$~eV, when deflections by
magnetic fields become more important, a clustered component still
exists in the AGASA, Yakutsk and HiRes stereo data, but with reduced
significance.

The small-scale clustering of UHECR arrival directions has been discussed by 
various authors. These works can be divided into two main
groups: The first one calculates the significance of the small-scale
clusters~\cite{Uchihori:1999gu,Goldberg:2000zq,Burgett:2003yg,Harari:2004py,Yoshiguchi:2004np},
while the second group of works uses the data to estimate parameters like the
density $n_s$ of sources or the strength of magnetic
fields~\cite{Wa96,Dubovsky:2000gv,Fodor:2000yi,Yoshiguchi:2002rb,Yoshiguchi:2003vs,Blasi:2003vx}. 
The authors of Ref.~\cite{Wa96} pointed out, to our knowledge for the
first time, that the observation of small-scale clusters allows to
determine the number density of CR sources. In practice, the observed
small-scale clusters of AGASA  were used to estimate the number CR
sources first in the pioneering work of Dubovsky, Tinyakov and
Tkachev~\cite{Dubovsky:2000gv}. 
Previous analyses of the significance of the small-scale clusters
observed by AGASA used a continuous distribution of sources as
a test hypothesis. Such a distribution has the advantage of
being model-independent and gives a lower limit on the significance
of clustering for a finite number of sources,
as long as deflections by magnetic fields can be neglected. 
Here, we investigate the significance of
the small-scale clusters within a realistic model of
UHECR protons propagating from astrophysical sources distributed
uniformly in the Universe. In particular, we calculate the number
of true clusters, i.e. those with CRs from the same source, as
function of $n_s$. We show that the asymptotic limit of a continuous
distribution of sources is 
reached only for an unrealistic high density of sources, $n_s \gg
10^{-2}/{\rm Mpc}^3$, where the latter value corresponds  
to one source per galaxy. 
We estimate also the number density of CR sources assuming small
deflections of CRs in galactic and extragalactic magnetic fields. Our
Monte Carlo procedure is very similar to the one of
Ref.~\cite{Blasi:2003vx}. We derive however confidence levels for the
consistency of arbitrary source densities with the clustering observed 
by the AGASA experiment, while Ref.~\cite{Blasi:2003vx} considered
exemplary only three values for the source density. Moreover,
our analysis shows strong deviations from Gaussianity for the
probability distribution of the autocorrelation function.
For the best-fit value $n_s=(1$--$4)\times 10^{-5}/{\rm Mpc}^3$
derived from the AGASA data, the probability that at least one observed
cluster is from a true point source is larger than $99.97\%$.
For such densities, we predict that the Pierre Auger Observatory (PAO)
\cite{PAO} will be able to determine $n_s$ within a factor of ten at 
$2\,\sigma$~C.L. after one year of data taking. In the same time, 
the PAO will establish that clustering is not by chance at the
at the $5~\sigma$ level for any {\em estimated\/} source density
smaller than $n_s=10^{-4}/$ Mpc$^3$.

\section{Analysis of the AGASA and HIRES data}
The authors of Ref.~\cite{tt_point} used first the angular two-point
auto-correlation function $w$ discussing the significance of
small-scale clustering in the arrival directions of UHECRs. Since the 
signal of point-like sources should be concentrated around
$\ell_{ij}=0$, we restrict our analysis to the value of $w$ in the first bin,  
\be
 w_1 = \sum_{i=1}^N\sum_{j=1}^{i-1} \Theta(\ell_1-\ell_{ij}) \,,
\ee
where $\ell_{ij}$ is the angular distance between the two cosmic rays
$i$ and $j$, $\ell_1$ the chosen bin size, $\Theta$ the step function,
and $N$ the number of CRs considered.

A draw-back of using only the first bin of the autocorrelation function
$w$ is the dependence of the results on $\ell_1$. As a possible
solution, one can perform a scan over different bin sizes and
calculate the resulting penalty factor~\cite{Finley:2003ur}. 
However, the result then still depends on the minimal and maximal bin
size used in the scan: Choosing the scan range too large reduces the
signal-to-noise ratio and thus diminishes the signal, while a too
small range overestimates the signal.
Following a different approach, we generated
artificial data sets from a single point source, deflected them in the
magnetic field, and finally smeared their arrival directions 
according to the angular resolution of the experiment. 
Then we chose the best binning size $\ell_1$ such that
the probability to observe an experimental value $\hat w_1^\ast$ by
chance is minimized as function of $\ell_1$. Here, $\hat w$
is the normalized auto-correlation function,
\be
 \hat w =
 \frac{2\Omega_{\rm exp}}{{\Omega_{\rm bin}}N(N-1)} \: w \,,
\ee
where $\Omega_{\rm exp}$ and $\Omega_{\rm bin}$ denote the solid angle 
with non-zero exposure of the experiment and of the bin considered,
respectively. 
Without the effect of the Galactic magnetic
field, the optimal value for $\ell_1$ found e.g. for the angular
resolution of the AGASA experiment is $\ell_1\approx 2^\circ$; 
including the effect of the Galactic magnetic field we found as optimal
range of values  $\ell_1\approx 2-4^\circ$.
Similar as for $\ell_1$, we could try to find  the optimal minimal energy
$E_{\min}$ of events taken into account. Earlier analyses found as
penalty factor for the scan over $E_{\min}$ in the AGASA data only a factor
three~\cite{tt_point,Finley:2003ur}. 

\begin{figure}
\epsfig{file=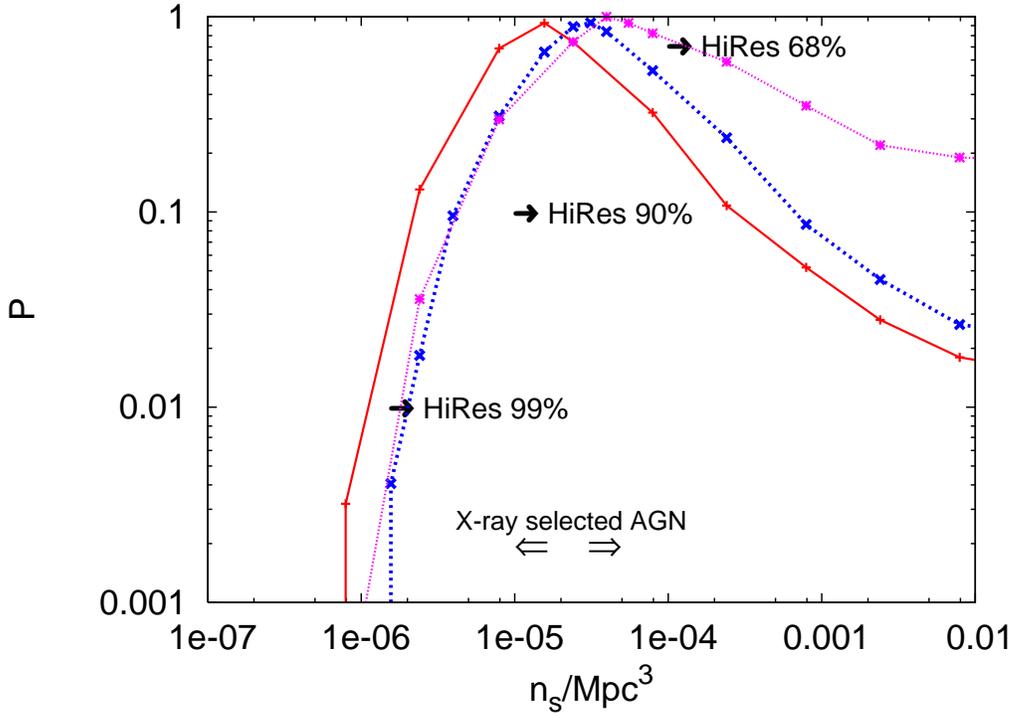,width=0.7\textwidth,angle=270}
\caption{Consistency level $P$ of an uniform source
  distribution with the AGASA data as function of the density $n_s$,
  with $N=57$ (solid line) and $N=72$ (dotted line) events for
  $\ell_1=2.5^\circ$, and with $N=57$ (dashed line) for $\ell_1=5^\circ$.
  Lower limits on $n_s$ from the HiRes stereo data (arrows) and 
  the  density range of $X$-ray selected AGNs with $X$-ray luminosity $L>
  10^{43}$~erg/s are also shown. 
\label{agasa_n}}
\end{figure}

We generate sources with constant comoving density $n_s$ up to the maximal
redshift $z=0.2$; we have checked that the flux of sources further
away is  negligible 
above $4\times 10^{19}$~eV. Then we choose a
source $i$ with equatorial coordinates R.A. and $\delta$
at comoving radial distance $R_i$ according to the declination
dependent exposure of the experiment and the weight
\be
 g_i = \frac{1+z_{\min}}{1+z_i}\left(\frac{R_{\min}}{R_i}\right)^2  \,.
\ee
Here, $R_{\min}$ and $z_{\min}$ are the distance and redshift of the
nearest source in the sample, respectively, and we have assumed the
same luminosity for all sources. Then CRs are generated according to
the injection spectrum $dN/dE\propto E^{-\alpha}$, where we fix 
$\alpha=2.7$ to reproduce best the AGASA energy spectrum below the GZK
cutoff. We propagate CRs until their energy is below
$E_{\min}$ or they reach the Earth. In that case, we take into
account the energy-dependent angular deflection through the Galactic
magnetic field and the angular resolution of the experiment.
For the angular resolution, we use a spherical
Gaussian density $\propto \exp(-\ell^2/(2\sigma_l^2))\sin(\ell)d\ell$
with $\sigma_\ell/{\rm degree} = \max(0.8,-0.6\log(E/{\rm eV})+13)$ 
for AGASA and the PAO, and $\sigma_\ell/{\rm degree} = 0.4$ for HiRes,
respectively, and for the Galactic magnetic field we use a shift by 
$\ell/{\rm degree}= 1.6\times 10^{20}~{\rm eV}/(E\cos({\rm R.A.})$.

The basic outcome of a sample of Monte Carlo simulations for fixed
parameters $\theta=n_s,\ell_1,\ldots,$ is a binned distribution
$p(w_1|\theta)$ for the 
values $w_1$ of the auto-correlation function. With how much
confidence can we accept or reject the hypothesis that the
experimentally measured value $w_1^\ast$ is drawn from $p(w_1|\theta)$?
In previous analyses, the test hypothesis was a continuous, isotropic
distribution of sources on a sphere $S^2$ for which one expects lower values
of $w_1$ than measured. Therefore, the probability that $w_1^\ast$ is
consistent with $p(w_1;\theta)$ was calculated as
$P_>(w_1^\ast,S^2) =
 \sum_i  p_i(w_1|S^2) \Theta \left( w_1-w_1^\ast \right)$.
This asymmetric definition fails when one wants to reject
both cases with too much and with too little clustering. 
We shall use as a more symmetric measure for the discrepancy
between $w_1^\ast$ and $p(w_1|\theta)$ the area between the measured
value $w_1^\ast$
and the median $w_{1/2}$ of the distribution $p(w_1|\theta)$.

In Fig.~\ref{agasa_n}, we show $P(w_1^\ast,\theta)$ as function of
$n_s$ for three different cases: The publicly available AGASA data set
until May 2000~\cite{up-date} ($N=57$ or $E_{\min}=4\times
10^{19}$~eV) for the two bin sizes $\ell_1=2.5^\circ$ ($w_1^\ast=7$)
and $\ell_1=5^\circ$ ($w_1^\ast=10$), and the complete AGASA data
set~\cite{c} ($N=72$ or $E_{\min}=4\times 10^{19}$~eV, bin size
$\ell_1=2.5^\circ$ and $w_1^\ast=8$). 
Remarkably, the most likely value for the source density, 
$n_s=(1$--$4)\times 10^{-5}/{\rm Mpc}^3$ is stable against an increase
of the data set and a change in bin size. 
A similar value for $n_s$ was found previously by the authors of 
Ref.~\cite{Blasi:2003vx}, while earlier 
analyses~\cite{Dubovsky:2000gv,Fodor:2000yi}
using only events above $E=10^{20}$~eV obtained larger values for $n_s$.
The steep decrease of $P(w_1^\ast,\theta)$ for low $n_s$
excludes already now uniformly distributed sources with
much lower density than $10^{-6}/{\rm Mpc}^3$. 
For comparison, we show also the estimated density
of powerful AGNs with $X$-ray luminosity $L> 10^{43}$~erg/s in the
energy range 0.2--5~keV, 
$n_s\sim (1-5)\times 10^{-5}/$Mpc$^3$~\cite{Steffen:2003uv}. 
The  density of Seyfert galaxies is about a factor of 20 higher.
Note that most often only very specific subsets of AGNs with much lower
densities are discussed as sources of UHECRs. On the other
side, $P(w_1^\ast|\theta)$ decreases only slowly for large $n_s$. With
the present AGASA data set it is therefore difficult to exclude large
source densities.

Recently, the HiRes collaboration published an analysis of their stereo
data~\cite{HiRes-stereo}. Their data set with $N=27$ events above 
$4\times 10^{19}$~eV contains no doublet within $\ell_1=2.5^\circ$ and
$5^\circ$~\cite{chad}. Therefore, the HiRes data alone are consistent
with a continuous source distribution. But since the
number of events is small and $p(w_1|\theta)$ is a broad distribution, 
the HiRes data are also consistent with the best-fit value for $n_s$
from the AGASA data, at $53\%$ and $21\%$ C.L. for  
$\ell_1=2.5^\circ$ and $5^\circ$, respectively.
In Fig.~\ref{agasa_n}, we show also lower limits on $n_s$ for
$\ell_1=5^\circ$ from the HiRes stereo data. Similar conclusions were   
recently obtained in Ref.~\cite{Yoshiguchi:2004np}. 
The HiRes data favor a larger value of $n_s$ than AGASA and may
indicate that practically all Seyfert galaxies contribute to the CR
flux above $4\times 10^{19}$~eV.

The effect of extragalactic magnetic fields on the above results is
negligible, if the deflection is $2^\circ$ on 500~Mpc
propagation distance as found for a large part of the sky in
Ref.~\cite{Dolag:2003ra}.  
The assumption of equal luminosity of
all sources gives a lower bound on the possible number of sources
\cite{Dubovsky:2000gv}.  A large additional population of faint
sources cannot be excluded, if their contribution to the UHECR flux is
sufficiently small. However, it is unlikely that any large population of
sources can accelerate CRs to energies $\gsim 10^{19}$~eV.

\begin{figure}
\epsfig{file=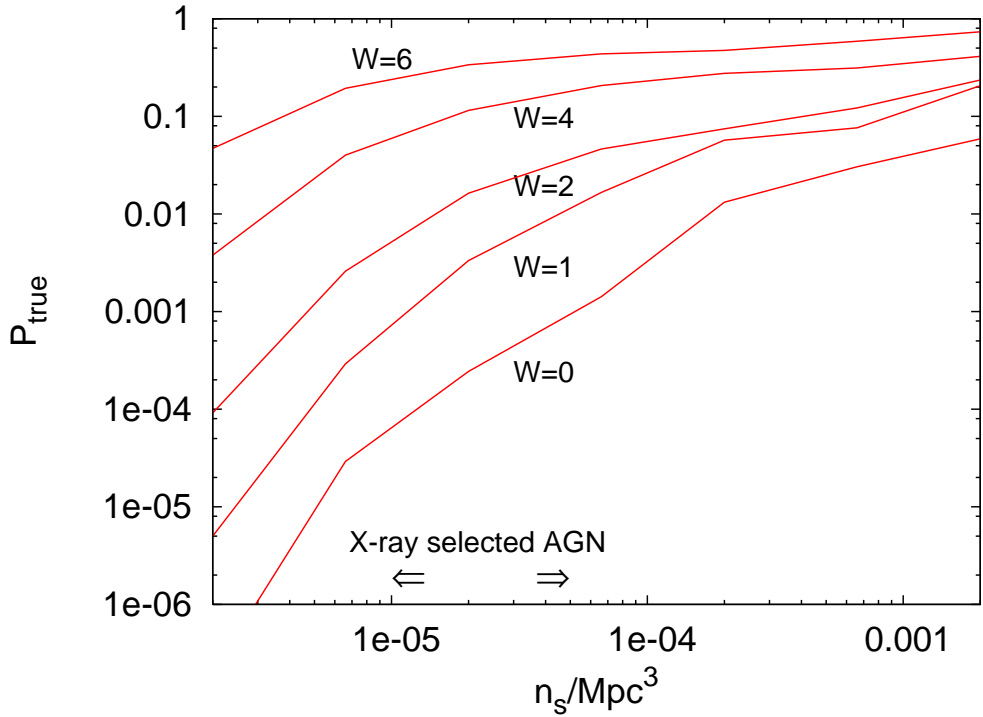,width=0.7\textwidth,angle=270}
\caption{Probability $P_{\rm true}$  to have a smaller or equal value of the
  source auto-correlation function $W$ as function of the source
  density $n_s$ for  $\ell_1=2.5^\circ$, 57 events and $w_1^\ast=7$. 
  The line $W=0$ corresponds to the case that all observed clusters are
  by chance.
  The  density range of $X$-ray selected AGNs with $X$-ray luminosity $L>
  10^{43}$~erg/s is also shown.
\label{agasa_true}}
\end{figure}

Apart from the auto-correlation function $w_1$ of the observed arrival
directions of CRs, i.e. including deflections and the finite
experimental resolution, we can calculate also the 
auto-correlation function of the sources,
$W=\sum_{i=1}^N\sum_{j=1}^{i-1} \delta_{ij}$, 
with $\delta_{ij}=1$ when the two CRs are from the same source
and $\delta_{ij}=0$ otherwise. 
Using only simulations which reproduce the
observed value $w_1^\ast$ defines $p(W|w_1^\ast,n_s)$.
In Fig.~\ref{agasa_true}, we show the probability $P_{\rm true}$ 
to have a value of the auto-correlation function smaller or equal than
$W$, $P_{\rm true}(W) =
\sum_{W^\prime\leq W}  p(W^\prime |w_1^\ast,n_s)$ 
as function of $n_s$ for $N=57$ events, $w_1^\ast=7$ and
$\ell_1=2.5^\circ$. Since the difference between the emitted and the
observed direction of the CRs can be larger than $\ell_1$, the values
of $W$ can exceed $w_1^\ast$ for finite $n_s$.

The asymptotic behavior of $P_{\rm true}(W)$ is easily understandable:
For a single source, i.e. $n_s\to 0$, $P_{\rm true}(W)\to 0$ and
$W=N(N-1)/2$ for $N$ observed events. On the other hand, for
$n_s\to\infty$ all clusters are by chance and thus $P_{\rm true}(W)\to 1$.  
A priori, it is unclear if for source densities typical for, e.g.,
AGNs the distribution $P_{\rm true}$ is still
close to its limiting value for $n_s\to\infty$ or already strongly
changed.  We find that the limit $P_{\rm true}(W)\to 1$ is
approached only for unrealistic high 
source densities, $n_s\gg 10^{-2}/{\rm Mpc}^3$, where the latter
value corresponds to the density of ordinary galaxies. 
For smaller $n_s$, the probability
that at least one cluster observed by AGASA is real increases very fast 
and reaches 99.97\% at $n_s = 2\times 10^{-5}/{\rm Mpc}^3$. 
But even for densities as large as $n_s=10^{-3}/{\rm Mpc}^3$, less
than half of  the cluster are on average by chance.

\section{Prediction for the PAO}
The value of the auto-correlation function $w_1$ is dominated by
clusters with high multiplicities and, thus suffers from large cosmic
variance. Moreover, the overlap of the distributions 
$p(\hat w_1|\theta)$ for different $n_s$ will be only very slowly
reduced by collecting more data. We have found that 
the fraction of singlets is a more stable quantity
against cosmic variance: singlet 
events can come from larger distances than multiplets and are thus
less affected by variations of the source distributions.
Therefore we propose to use the distribution $p(N_1|\theta)$ of the
number of singlet events instead of $p(\hat w_1|\theta)$ to estimate
$n_s$ from the PAO data.

\begin{figure}
\epsfig{file=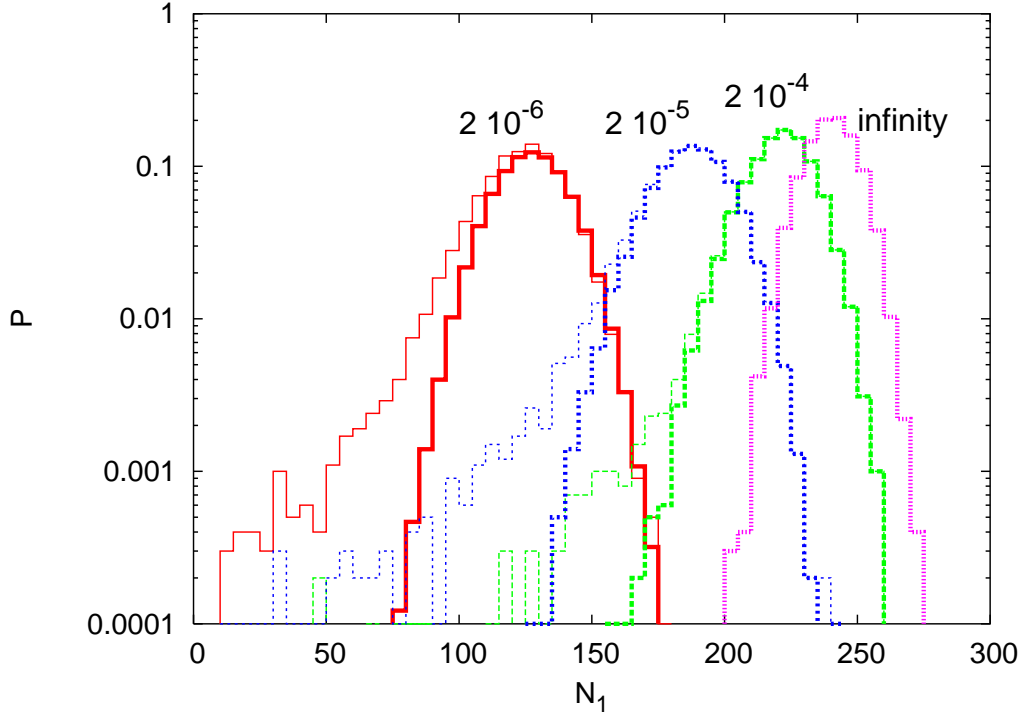,width=0.7\textwidth,,angle=270}
\caption{Predicted distributions $p(N_1|\theta)$ for source densities
  $n_s=10^{-6}$, $10^{-5}$, $10^{-4}/{\rm Mpc}^3$ and for continuous source
  distribution after one year data taking of the PAO ($N=300$). 
\label{auger_sing}}
\end{figure}

\begin{figure}
\epsfig{file=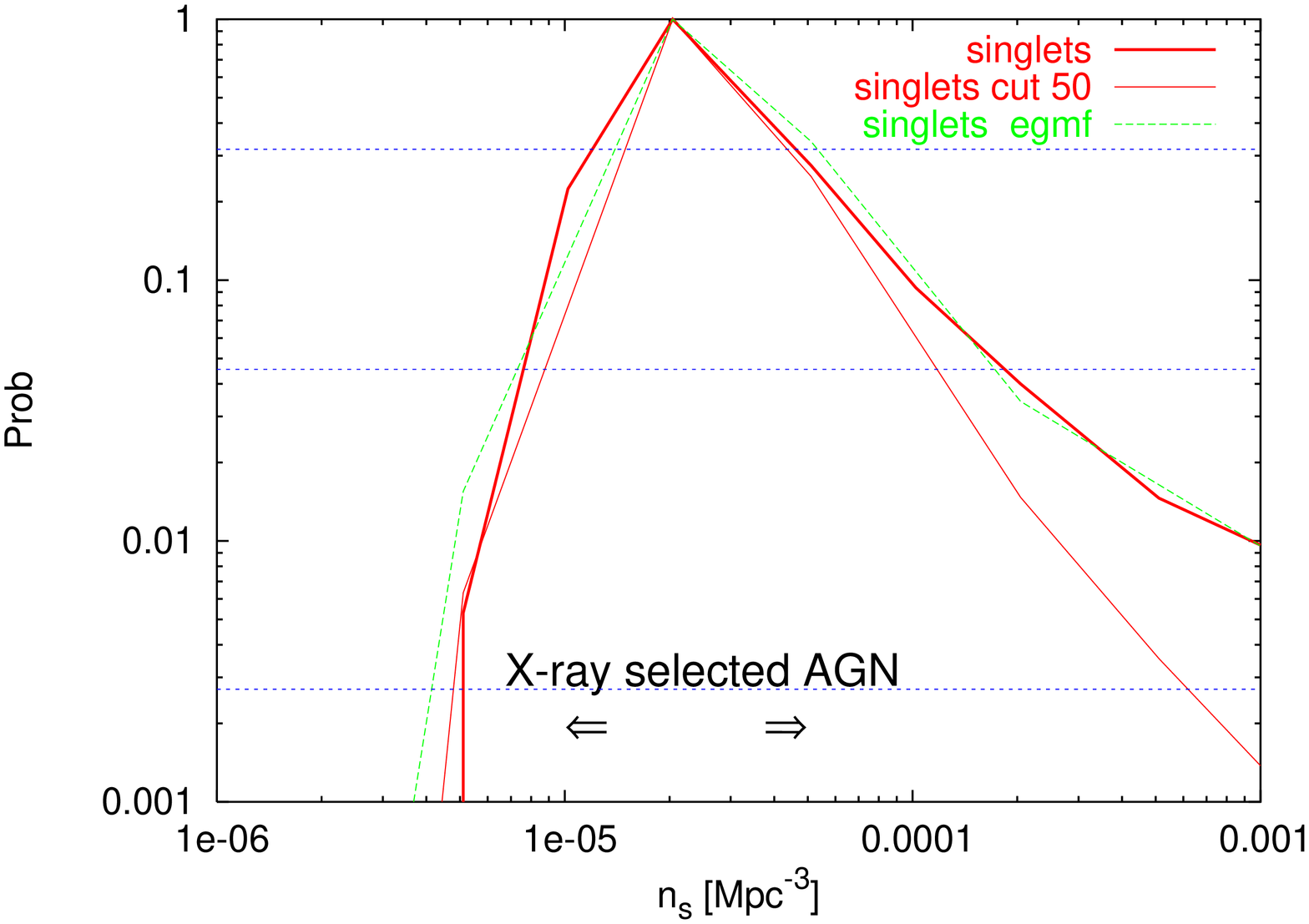,width=0.7\textwidth,,angle=270}
\caption{Confidence level for the estimate of the source density $n_s$ from
  the distribution $p(N_1|\theta)$; without and with multiplicity
  cut, and with extragalactic magnetic field ($N=300$ events). 
  The  density range of $X$-ray selected AGNs with $X$-ray luminosity $L>
  10^{43}$~erg/s is also shown.
\label{auger_ns}}
\end{figure}

In Fig.~\ref{auger_sing}, we show the distributions 
$p(N_1|\theta)$ for $N=300$ events expected in one  
year running of the PAO for $n_s=10^{-6}$,
$10^{-5}$ and $10^{-4}/{\rm Mpc}^3$, together with the one for the
limit $n_s\to\infty$.
The distributions $p(N_1|\theta)$ are
characterized by power-law tails towards small $N_1$, and these tails
seem to prevent a clear separation of different densities. 
However, a very small fraction of singlet events is caused in most
cases by a single nearby source producing 
clusters with very high multiplicity. We eliminate these exceptional
cases by considering only samples where the highest
multiplet is a 50-plet for $N=300$. This would
corresponds to 10 events from a single source in the case of AGASA.
If indeed an exceptionally bright source would be found by the PAO, 
the region around this source should be excluded from the analysis.
As an example, we show in Fig.~\ref{auger_ns} for $n_s=2\times
10^{-5}/{\rm Mpc}^3$ the probability with which the PAO estimates 
the value of $n_s$. The elimination of clusters with
very high multiplicity reduces the cosmic variance and thereby
increases the precision of the estimate for $n_s$. 
The influence of extragalactic magnetic fields on $p(N_1|\theta)$ is
clearly negligible, if the average deflection with $\approx 2^\circ$
on 500~Mpc is as small as found in Ref.~\cite{Dolag:2003ra}. If the
average deflection is closer to the values found in
Ref.~\cite{ems}, the average number of singlet events for a fixed
number of sources would increase and become more and more
indistinguishable from the case of an infinite number of sources.

Finally, we want to estimate how well the PAO can establish that the
clustering is not by chance, or equivalently, that the number of
sources is finite. From an experimental point of view, the PAO will measure
a certain value $w_1^\ast$ of the auto-correlation function. From this
measurement. one can estimate the density $n_s$ of sources.
For $N=300$ and $n_s=2\times10^{-5}$ the mean of $p(w_1|\theta)$ is 
$\langle w_1\rangle=138$. On the other hand, the largest
value found in $10^6$ simulations for a continuous distributions is
$w_1=67$.  Thus for any {\em estimated\/} source density smaller than 
$n_s=10^{-4}/$Mpc$^3$ the PAO can establish clustering with chance 
probability smaller than $10^{-6}$. The smallest value $w_1$ compatible at
99\% C.L. with a true density $n_s=2\times10^{-5}/$Mpc$^3$ is only at
$0.1\%$ compatible with an infinite number of sources.

\section{Summary}
We have investigated the significance of the small-scale clustering of
the arrival directions of UHECRs assuming a finite number of 
uniformly distributed proton sources 
and small deflection of CRs in extragalactic magnetic fields.
The AGASA data favor as source density $n_s\sim 10^{-5}/$Mpc$^{3}$, 
a value where the probability that at least one observed
cluster is from a true point source is larger than $99.97\%$.
Even for densities as large as $n_s=10^{-3}/{\rm Mpc}^3$, less than
half of the cluster are on average by chance.

At present, the sparse AGASA data set cannot exclude firmly that the
clustering is by chance without the prior knowledge of the source
density. In contrast, the PAO will confirm clustering from a finite
number of point sources within one year of data taking at the 
the $5~\sigma$ level for any source density $n_s<10^{-4}$~Mpc$^{-3}$. 
The PAO will also measure the density of UHECR sources within a factor
of a few,
and check the assumption of uniformly distributed sources. If the PAO
detects no significant clustering, then two possible explanations are that
the extragalactic magnetic fields are, especially in the voids, larger
than expected or that the UHECR primaries are nuclei.

\section*{Acknowledgments}
We are grateful to  V.~Berezinsky, R.~Cousins, C.~Finley, G.~Raffelt and
M.~Teshima for helpful discussions. 
M.K.\ acknowledges support by an Emmy-Noether grant of the DFG, 
and D.S.\ was supported in part by the NASA ATP grant NAG5-13399. 
D.S.\ thanks the MPI for hospitality during part of this work.


\end{document}